\documentclass[a4paper,11pt]{article}
\usepackage{pos}
\newcommand{\GeV}{{\ensuremath\rm GeV}}
\newcommand{\TeV}{{\ensuremath\rm TeV}}

\newcommand{\fb}{{\ensuremath\rm fb}}
\newcommand{\pb}{{\ensuremath\rm pb}}
\newcommand{\al}{\alpha}
\newcommand{\be}{\beta}

\title{Di-Higgs production in BSM models}

\author*[a]{Tania Robens}

\affiliation[a]{Rudjer Boskovic Institute\\
  Bijenicka cesta 54, 10000 Zagreb, Croatia}

\emailAdd{trobens@irb.hr}

\abstract{I give a short overview on di-Higgs production in models that extend the Standard Model of particle physics by additional fields and particle content, including EFT prescriptions.\\ RBI-ThPhys-2022-34}

\FullConference{%
  The Tenth Annual Conference on Large Hadron Collider Physics - LHCP2022\\
  16-20 May 2022\\
  online
}

\begin{document}
\maketitle

\section{Introduction}
After the discovery of a scalar boson that complies with the properties of the Standard Model (SM) Higgs boson \cite{ATLAS:2012yve,CMS:2012qbp}, particle physics has entered an exciting era. 
The confirmation that the scalar sector realized by nature indeed corresponds to the one predicted by the SM also requires an accurate knowledge of triple and quartic scalar couplings. Projections for the accuracy with which these couplings can be determined at future colliders are currently in the $\%$ range, from roughly $50\%$ accuracy at the HL-LHC up to projected $2-3\%$ at future high-energy lepton colliders (see e.g. \cite{DiMicco:2019ngk,Buttazzo:2020uzc,ILCInternationalDevelopmentTeam:2022izu,Dasu:2022nux}).

In this context, it is also interesting to investigate the modification of $h_{125} h_{125}$ rates by the presence of new states, either directly via a heavy resonance e.g. in the s-channel or via modified couplings in an effective field theory (EFT) approach, where the additional possible new heavy resonances have been integrated out.
For reference, the current value for di-Higgs production in the SM at a 13 \TeV~ LHC is given by $31.05\,\fb$ (see \cite{Grazzini:2018bsd,hhtwiki} for details and uncertainties).

\section{Resonance-enhanced di-scalar production}

\begin{figure}
\begin{center}
\includegraphics[width=0.45\textwidth]{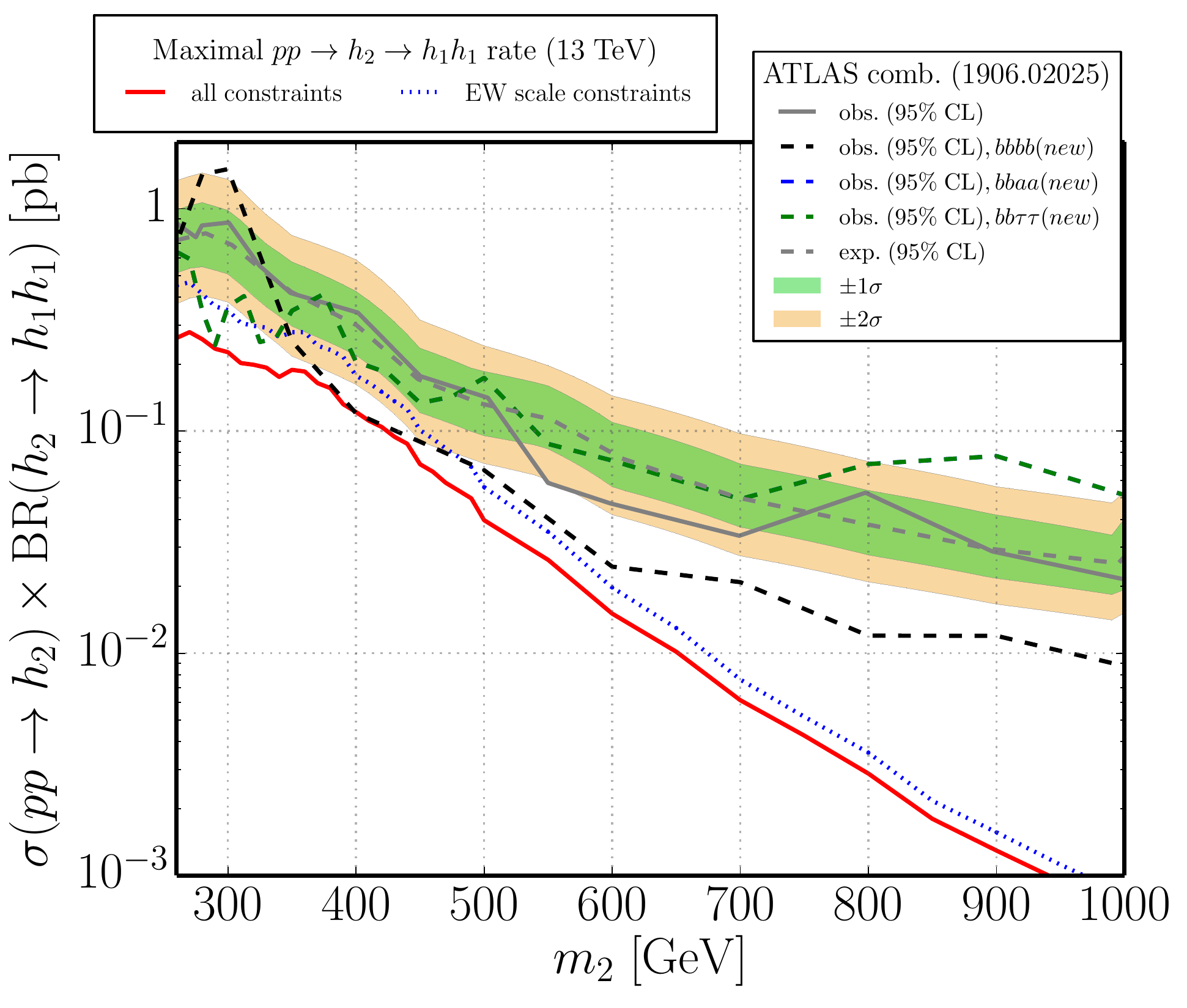}
\caption{\label{fig:singenh} Possible maximal bounds as well as current constraints for resonance-enhanced di-Higgs production in the $\mathbb{Z}_2$-symmetric singlet extension. Updated from \cite{Robens:2022oue}. Experimental constraints from \cite{ATLAS:2019qdc,ATLAS:2022hwc,ATLAS:2021ifb,ATLAS-CONF-2021-030}. }
\end{center}
\end{figure}

A simple example is the enhancement of the di-Higgs final state by mediation of an s-channel resonance, where the resonance corresponds to a second CP-even neutral scalar that mixes with the 125 \GeV~ particle. Imposing an additional symmetry on this model, the number of additional free parameters can be constrained to 3, which are typically chosen to be the second scalar mass $m_2$, a mixing angle $\sin\al$, and the ratio of the vacuum expectation values $\tan\be$. This model has e.g. been discussed at length in \cite{Robens:2015gla,Robens:2016xkb,Ilnicka:2018def,Robens:2022oue}. In figure \ref{fig:singenh}, I show the resonance enhanced rates for di-Higgs production in that model, taking current theoretical as well as experimental constraints into account. This figure includes the experimental constraints from ATLAS run I combination \cite{ATLAS:2019qdc} as well as dedicated searches with full run II data \cite{ATLAS:2022hwc,ATLAS:2021ifb,ATLAS-CONF-2021-030}\footnote{I thank D. Azevedo for providing exclusion limits for $b\,\bar{b}\tau^+\tau^-$ \cite{ATLAS-CONF-2021-030} in a digitized format, as used in \cite{Abouabid:2021yvw}.}, and corresponds to an update of figure 1 in \cite{Robens:2022oue}.

\begin{figure}
\begin{center}
\includegraphics[width=0.43\textwidth]{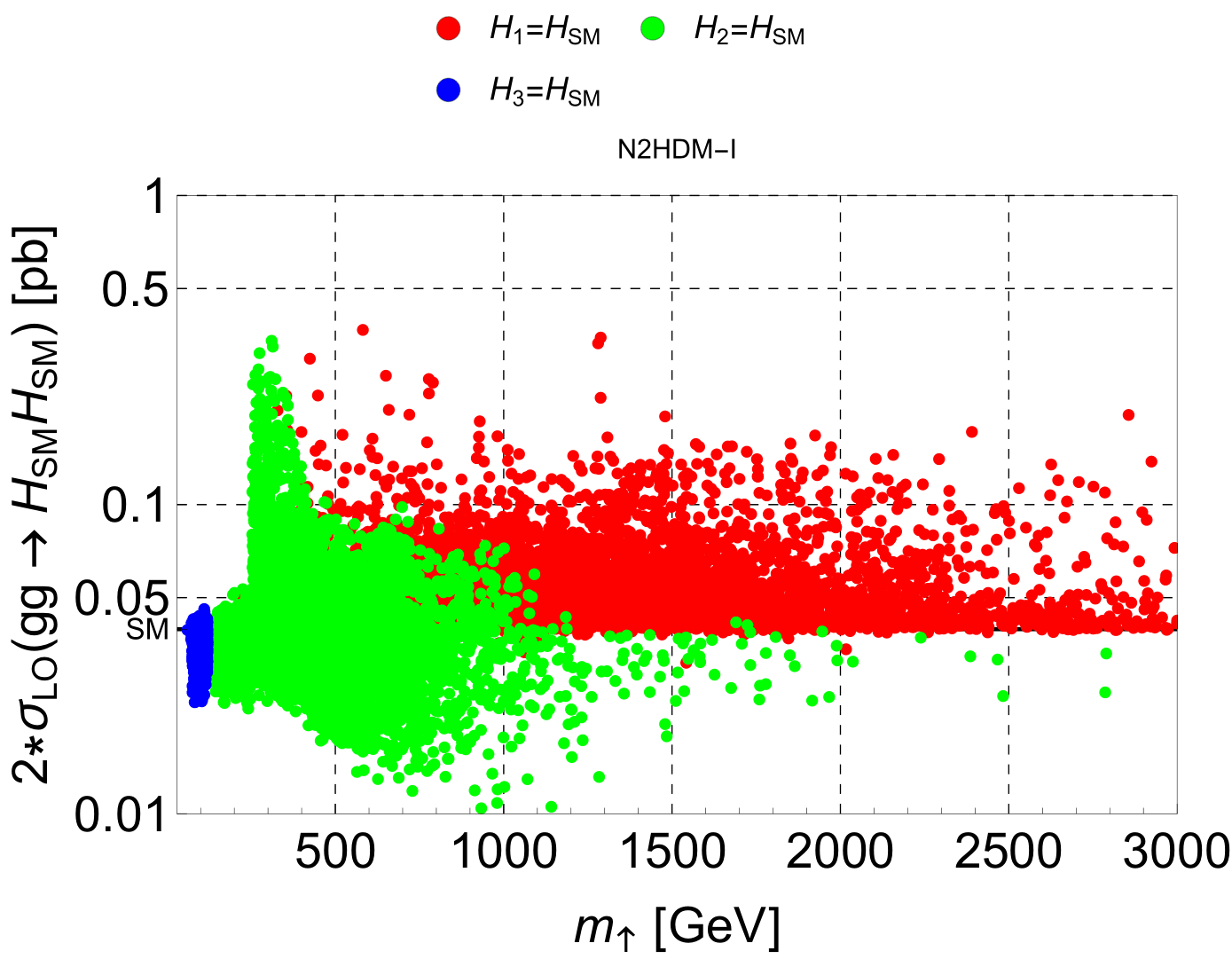}
\includegraphics[width=0.43\textwidth]{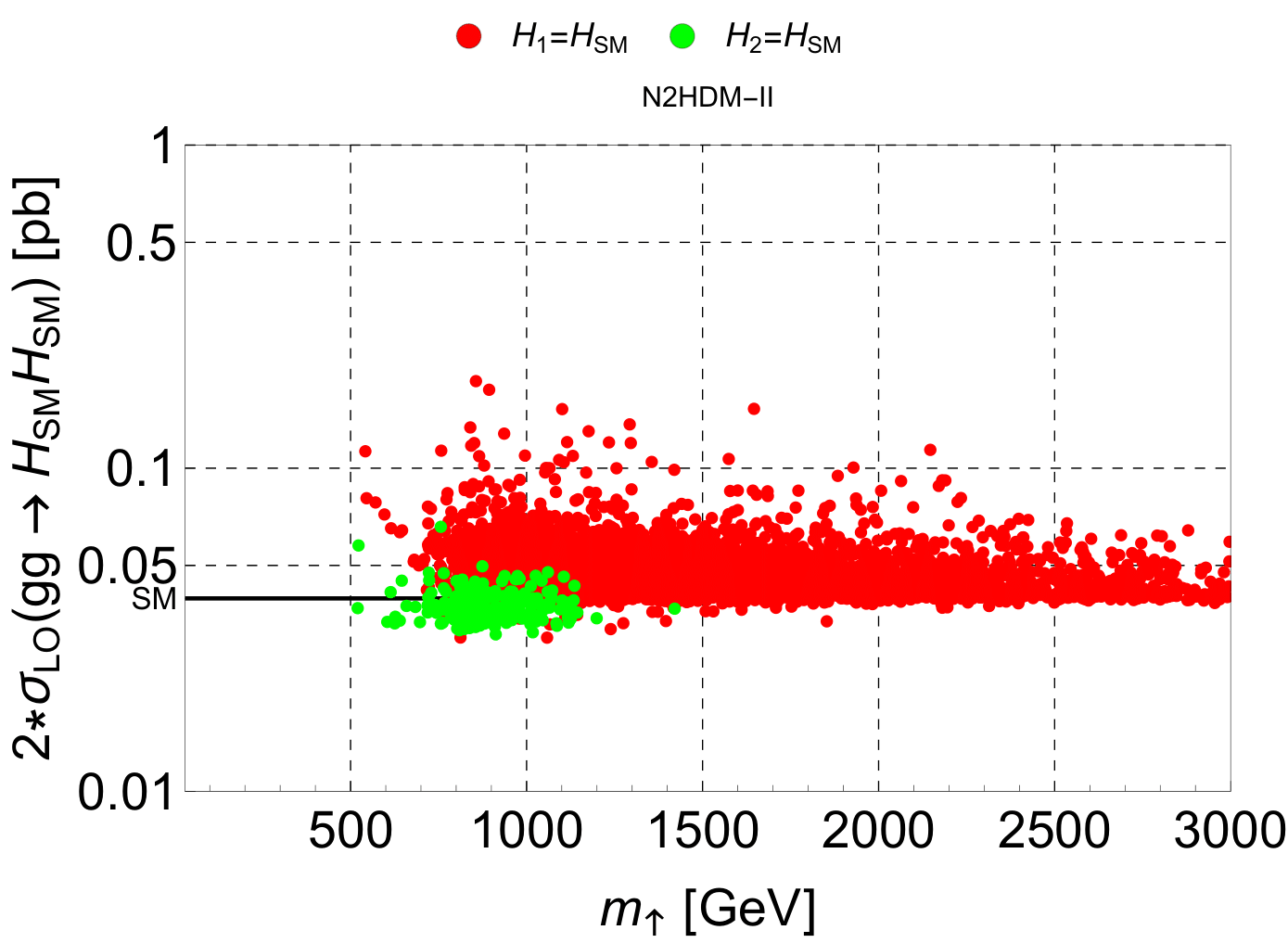}
\caption{\label{fig:maggieea} Rates for di-Higgs production in various scalar extended models. {\sl Top:} Real 2HDM, {\sl Bottom:} 
2HDM+singlet. {\sl Left:} Type I. {\sl Right:} Type II. 
The model features 3 CP-even neutral scalars; rates are shown as a function of the heaviest additional scalar. Taken from \cite{Abouabid:2021yvw}.}
\end{center}
\end{figure}

Apart from enhancement stemming from a relatively simple model, one can also consider di-Higgs rates that can be achieved in models with larger scalar sector extensions. Some examples in this direction have been presented in \cite{Abouabid:2021yvw}. 
In figure \ref{fig:maggieea}, I show two examples for the N2HDM, a two Higgs doublet model with an additional singlet. In this model 3 CP-even neutral scalar states exist. One of these has to comply with the current measurements of the 125 \GeV~ resonance at the LHC, while the others can take other mass values and couplings granted that all other theoretical and experimental constraints are fulfilled.
We see that in this model, the di-Higgs rate can be enhanced by roughly an order of magnitude with respect to the current SM prediction of around 30 \fb.
\section{EFT approach}
In effective field theory approaches, coupling modifiers are introduced that either multiply couplings already existing on the SM or introduce new coupling structures that do not exist in the SM. All corresponding production mechanisms can then contribute and also interfere. This leads to a very distinct behaviour in differential distributions, that depend on the specific values of these coupling modifiers. A prominent approach is then to define clusters in the corresponding parameter space that display similar behaviour. I here show an example of such clusters taken from \cite{Carvalho:2015ttv}
\footnote{One of the clusters presented in \cite{Carvalho:2015ttv} has been modified in \cite{Buchalla:2018yce}, where in addition higher-order corrections have been taken into account. I thank G. Heinrich for useful conversation regarding this point.}. Leading and next-to-leading order (NLO) production cross sections at the LHC can reach the \pb~ range and are given in \cite{CarvalhoAntunesDeOliveira:2130724,Buchalla:2018yce}, respectively. In \cite{Capozi:2019xsi}, an alternative set of clusters has been proposed based on NLO predictions.
\begin{figure}
\begin{center}
\includegraphics[width=0.8\textwidth]{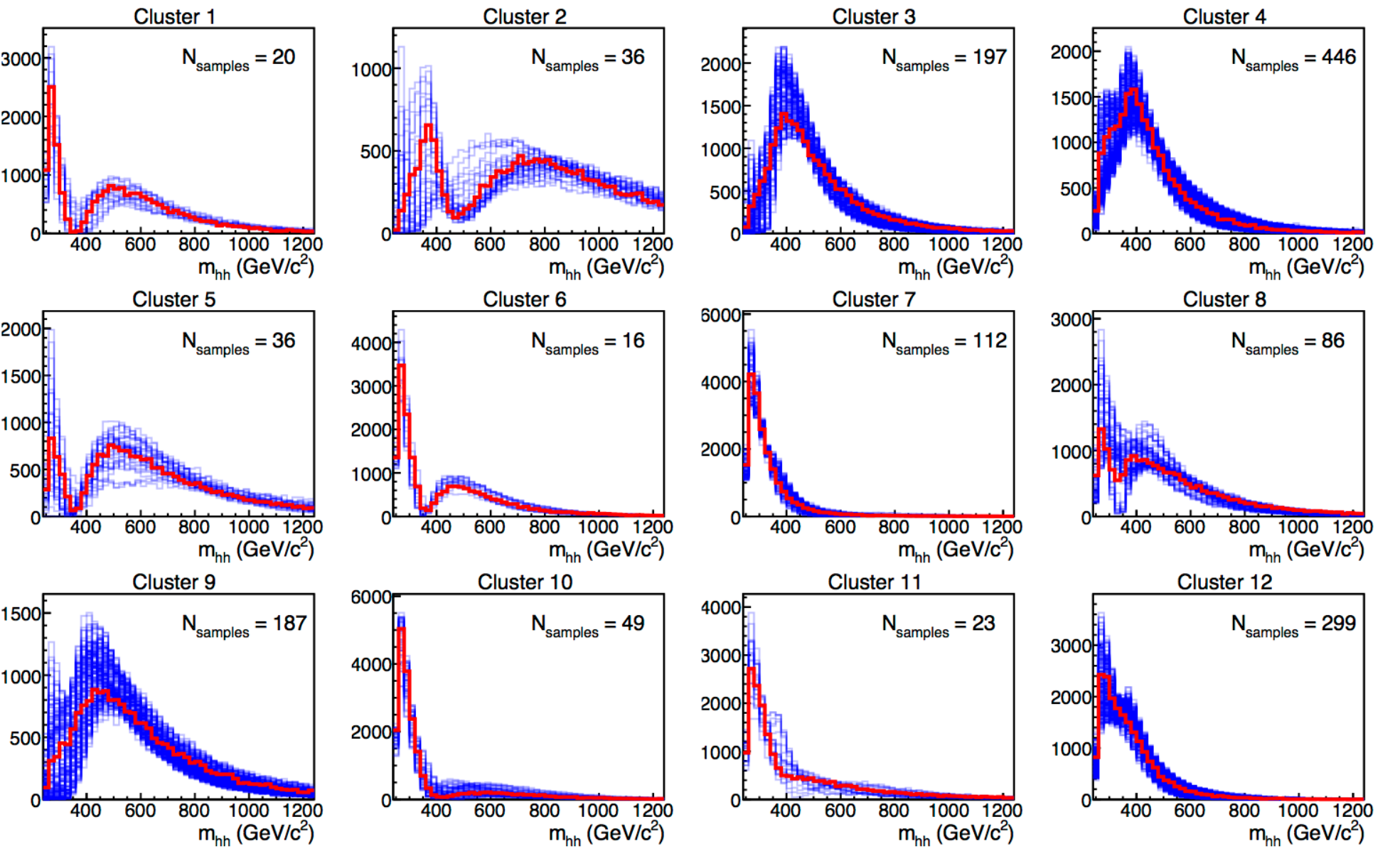}
\caption{Di-Higgs invariant mass distributions for specific clusters characterizing typical modifications if coupling parameters are altered from their SM values. More details in and figure from \cite{Carvalho:2015ttv}.}
\end{center}
\end{figure}
\section{Di-scalar states with masses $\neq\,125\,\GeV$}
I now present predictions for a model that features 3 CP-even neutral scalar states. The model extends the SM scalar sector by two real gauge singlet scalar fields and applies an additional $\mathbb{Z}_2\,\otimes\,\mathbb{Z}_2'$ symmetry 
\cite{Robens:2019kga,Robens:2022lkn}.
\begin{figure}
\begin{center}
\includegraphics[width=0.35\textwidth]{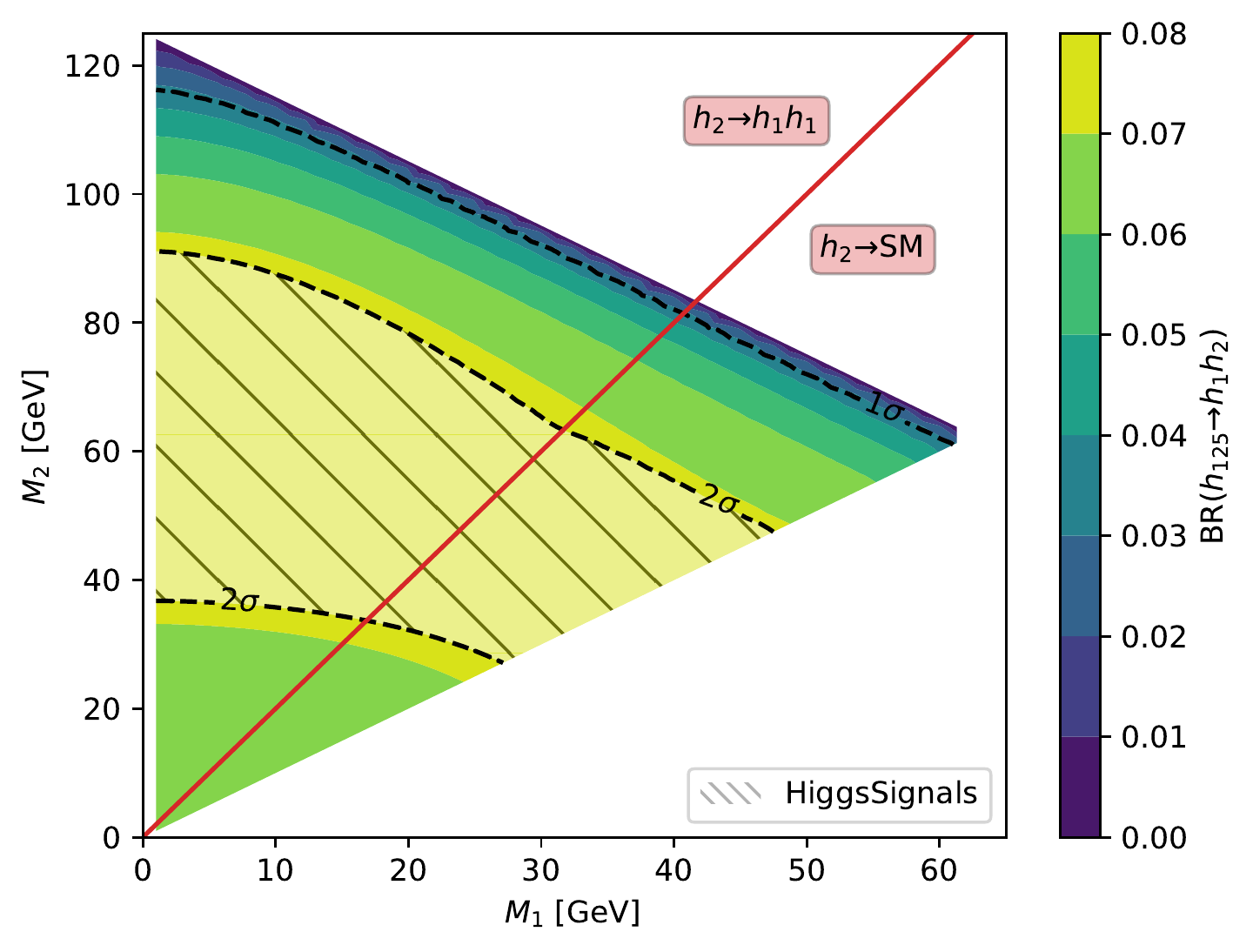}
\includegraphics[width=0.35\textwidth]{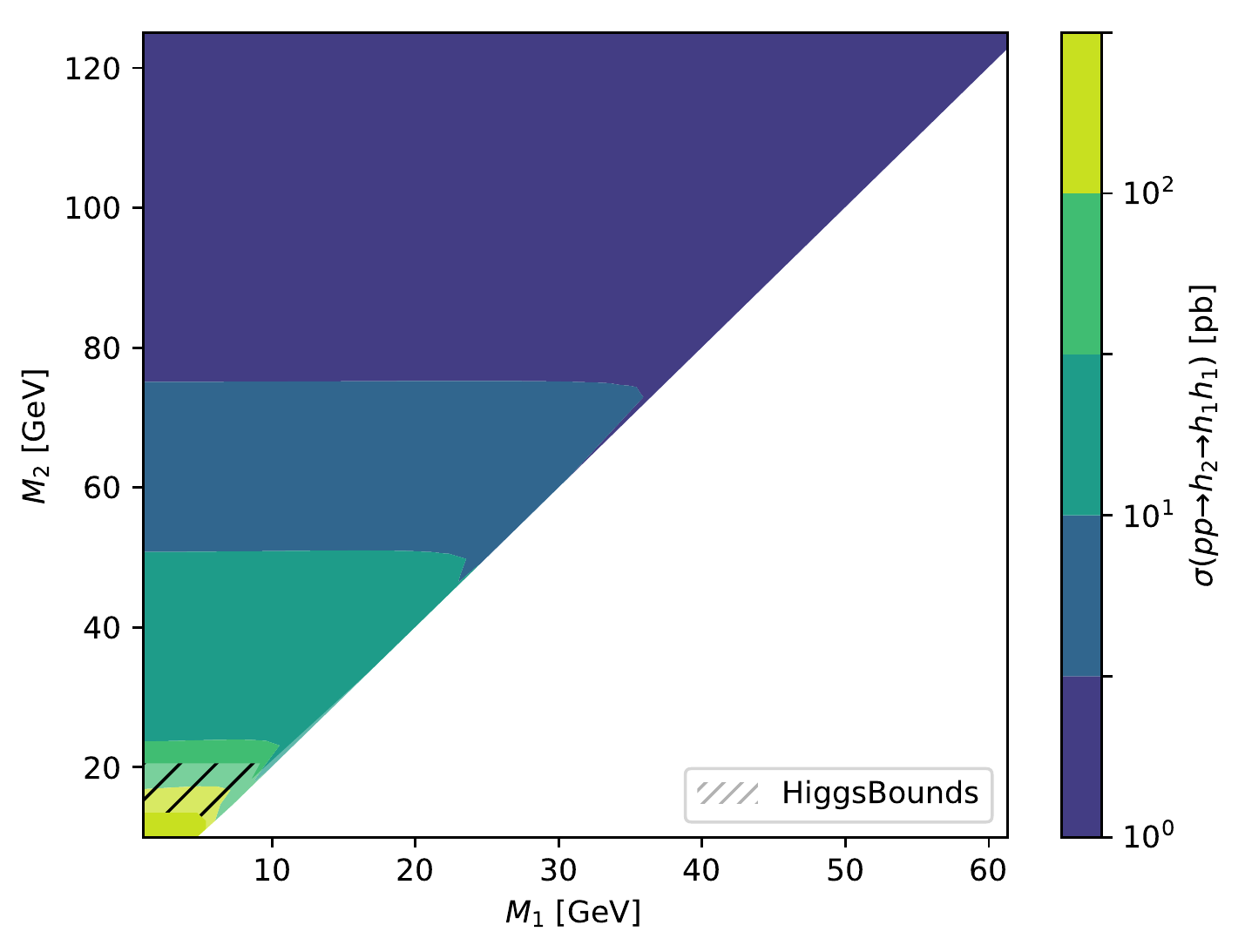}
\caption{\label{fig:trsm} Benchmark plane examples for cases where one or two masses differ from $h_{125}$. {\sl Left:} BP1, with $m_1, m_2\,\leq\,125\,\GeV$. Shown is the branching ratio $h_{125}\,\rightarrow\,h_1\,h_2$. {\sl Right:} BP4, with the same mass ranges. Shown is the factorized production time decay rate for $h_1\,h_1$ final states via $h_2$ mediation. Taken from \cite{Robens:2019kga}.}
\end{center}
\end{figure}
Figure \ref{fig:trsm} shows benchmark planes in this model for decays with of either the SM like scalar decaying into two lighter states with different masses (BP1, {\sl left}) or a light scalar decaying into two light scalars with the same mass (BP 4, {\sl right}), with cross sections ranging around 3.5 \pb~ and 60 \pb~ at a 13 \TeV~ LHC, respectively. In BP1, a $h_1 h_1 h_1$ final state is predominant if kinematically allowed. The light scalars decay mostly into $b\,\bar{b}$ in both scenarios, leading to $b\bar{b}b\bar{b}(b\bar{b})$ final states. Both scenarios are currently not yet explored by the LHC experiments.
\section{Discussion and outlook}
In this proceeding, I have discussed various new physics scenarios that allow to modify or enhance the SM prediction for di-Higgs final state rates. I have given a couple of examples, both for UV-complete models as well as in an EFT approach, and listed the corresponding references for further reading.


\begin{thebibliography}{10}

\bibitem{ATLAS:2012yve}
Georges Aad et~al.
\newblock {Observation of a new particle in the search for the Standard Model
  Higgs boson with the ATLAS detector at the LHC}.
\newblock {\em Phys. Lett. B}, 716:1--29, 2012, 1207.7214.

\bibitem{CMS:2012qbp}
Serguei Chatrchyan et~al.
\newblock {Observation of a New Boson at a Mass of 125 GeV with the CMS
  Experiment at the LHC}.
\newblock {\em Phys. Lett. B}, 716:30--61, 2012, 1207.7235.

\bibitem{DiMicco:2019ngk}
J.~Alison et~al.
\newblock {Higgs boson potential at colliders: Status and perspectives}.
\newblock {\em Rev. Phys.}, 5:100045, 2020, 1910.00012.

\bibitem{Buttazzo:2020uzc}
Dario Buttazzo, Roberto Franceschini, and Andrea Wulzer.
\newblock {Two Paths Towards Precision at a Very High Energy Lepton Collider}.
\newblock {\em JHEP}, 05:219, 2021, 2012.11555.

\bibitem{ILCInternationalDevelopmentTeam:2022izu}
Alexander Aryshev et~al.
\newblock {The International Linear Collider: Report to Snowmass 2021}.
\newblock 3 2022, 2203.07622.

\bibitem{Dasu:2022nux}
Sridhara Dasu et~al.
\newblock {Strategy for Understanding the Higgs Physics: The Cool Copper
  Collider}.
\newblock In {\em {2022 Snowmass Summer Study}}, 3 2022, 2203.07646.

\bibitem{Grazzini:2018bsd}
Massimiliano Grazzini, Gudrun Heinrich, Stephen Jones, Stefan Kallweit,
  Matthias Kerner, Jonas~M. Lindert, and Javier Mazzitelli.
\newblock {Higgs boson pair production at NNLO with top quark mass effects}.
\newblock {\em JHEP}, 05:059, 2018, 1803.02463.

\bibitem{hhtwiki}
https://twiki.cern.ch/twiki/bin/view/LHCPhysics/LHCHWGHH.

\bibitem{Robens:2015gla}
Tania Robens and Tim Stefaniak.
\newblock {Status of the Higgs Singlet Extension of the Standard Model after
  LHC Run 1}.
\newblock {\em Eur. Phys. J. C}, 75:104, 2015, 1501.02234.

\bibitem{Robens:2016xkb}
Tania Robens and Tim Stefaniak.
\newblock {LHC Benchmark Scenarios for the Real Higgs Singlet Extension of the
  Standard Model}.
\newblock {\em Eur. Phys. J. C}, 76(5):268, 2016, 1601.07880.

\bibitem{Ilnicka:2018def}
Agnieszka Ilnicka, Tania Robens, and Tim Stefaniak.
\newblock {Constraining Extended Scalar Sectors at the LHC and beyond}.
\newblock {\em Mod. Phys. Lett. A}, 33(10n11):1830007, 2018, 1803.03594.

\bibitem{Robens:2022oue}
Tania Robens.
\newblock {More Doublets and Singlets}.
\newblock In {\em {56th Rencontres de Moriond on Electroweak Interactions and
  Unified Theories}}, 5 2022, 2205.06295.

\bibitem{ATLAS:2019qdc}
Georges Aad et~al.
\newblock {Combination of searches for Higgs boson pairs in $pp$ collisions at
  $\sqrt{s} = $13 TeV with the ATLAS detector}.
\newblock {\em Phys. Lett. B}, 800:135103, 2020, 1906.02025.

\bibitem{ATLAS:2022hwc}
Georges Aad et~al.
\newblock {Search for resonant pair production of Higgs bosons in the
  $b\bar{b}b\bar{b}$ final state using $pp$ collisions at $\sqrt{s}$ = 13 TeV
  with the ATLAS detector}.
\newblock {\em Phys. Rev. D}, 105(9):092002, 2022, 2202.07288.

\bibitem{ATLAS:2021ifb}
Georges Aad et~al.
\newblock {Search for Higgs boson pair production in the two bottom quarks plus
  two photons final state in $pp$ collisions at $\sqrt{s}=13$ TeV with the
  ATLAS detector}.
\newblock {\em Phys. Rev. D}, 106(5):052001, 2022, 2112.11876.

\bibitem{ATLAS-CONF-2021-030}
{Search for resonant and non-resonant Higgs boson pair production in the $b\bar
  b\tau^+\tau^-$ decay channel using 13 TeV $pp$ collision data from the ATLAS
  detector}.
\newblock Technical report, CERN, Geneva, 2021.
\newblock All figures including auxiliary figures are available at
  https://atlas.web.cern.ch/Atlas/GROUPS/PHYSICS/CONFNOTES/ATLAS-CONF-2021-030.

\bibitem{Abouabid:2021yvw}
Hamza Abouabid, Abdesslam Arhrib, Duarte Azevedo, Jaouad~El Falaki, Pedro.~M.
  Ferreira, Margarete M\"uhlleitner, and Rui Santos.
\newblock {Benchmarking di-Higgs production in various extended Higgs sector
  models}.
\newblock {\em JHEP}, 09:011, 2022, 2112.12515.

\bibitem{Carvalho:2015ttv}
Alexandra Carvalho, Martino Dall'Osso, Tommaso Dorigo, Florian Goertz, Carlo~A.
  Gottardo, and Mia Tosi.
\newblock {Higgs Pair Production: Choosing Benchmarks With Cluster Analysis}.
\newblock {\em JHEP}, 04:126, 2016, 1507.02245.

\bibitem{Buchalla:2018yce}
G.~Buchalla, M.~Capozi, A.~Celis, G.~Heinrich, and L.~Scyboz.
\newblock {Higgs boson pair production in non-linear Effective Field Theory
  with full $m_t$-dependence at NLO QCD}.
\newblock {\em JHEP}, 09:057, 2018, 1806.05162.

\bibitem{CarvalhoAntunesDeOliveira:2130724}
Alexandra Carvalho Antunes De~Oliveira, Martino Dall'Osso, Pablo
  De~Castro~Manzano, Tommaso Dorigo, Florian Goertz, Maxime Gouzevitch, and Mia
  Tosi.
\newblock {Analytical parametrization and shape classification of anomalous HH
  production in EFT approach}.
\newblock 2016.

\bibitem{Capozi:2019xsi}
Matteo Capozi and Gudrun Heinrich.
\newblock {Exploring anomalous couplings in Higgs boson pair production through
  shape analysis}.
\newblock {\em JHEP}, 03:091, 2020, 1908.08923.

\bibitem{Robens:2019kga}
Tania Robens, Tim Stefaniak, and Jonas Wittbrodt.
\newblock {Two-real-scalar-singlet extension of the SM: LHC phenomenology and
  benchmark scenarios}.
\newblock {\em Eur. Phys. J. C}, 80(2):151, 2020, 1908.08554.

\bibitem{Robens:2022lkn}
Tania Robens.
\newblock {TRSM Benchmark Planes - Snowmass White Paper}.
\newblock In {\em {2022 Snowmass Summer Study}}, 5 2022, 2205.14486.

\end{thebibliography}

\end{document}